\PassOptionsToPackage{pdfpagelabels=false}{hyperref}
\documentclass[fleqn,usenatbib]{mnras}

\usepackage{newtxtext,newtxmath}       
\usepackage[T1]{fontenc}               
\usepackage{ae,aecompl}                
\usepackage{amsmath,mathtools,tensor}  
\usepackage{graphicx}                  
\usepackage{booktabs,multirow}         

\hypersetup{colorlinks=true,linkcolor=black,citecolor=black,filecolor=black,urlcolor=black}

\DeclarePairedDelimiter{\abs}{\lvert}{\rvert}
\DeclarePairedDelimiter{\paren}{\lparen}{\rparen}

\newcommand{\dd}{\mathrm{d}}

\newcommand{\mdot}{\dot{M}}
\newcommand{\rhor}{r_\mathrm{hor}}
\newcommand{\ugas}{u_\mathrm{gas}}
\newcommand{\pgas}{p_\mathrm{gas}}
\newcommand{\pmag}{p_\mathrm{mag}}
\newcommand{\tem}{T_\mathrm{EM}}
\newcommand{\tkin}{T_\mathrm{kin}}
\newcommand{\ten}{T_\mathrm{en}}
\newcommand{\tmass}{T_\mathrm{mass}}
\newcommand{\pjet}{P_\mathrm{jet}}
\newcommand{\vjet}{v^r_\mathrm{jet}}
\newcommand{\tp}{T_\mathrm{p}}
\newcommand{\te}{T_\mathrm{e}}
\newcommand{\rhigh}{R_\mathrm{high}}
\newcommand{\rlow}{R_\mathrm{low}}
\newcommand{\mdotedd}{\dot{M}_\mathrm{Edd}}
\newcommand{\mproton}{m_\mathrm{p}}
\newcommand{\sigmat}{\sigma_\mathrm{T}}

\newcommand{\ghz}{\mathrm{GHz}}
\newcommand{\msun}{M_\odot}
\newcommand{\mpc}{\mathrm{Mpc}}
\newcommand{\jy}{\mathrm{Jy}}
\newcommand{\yr}{\mathrm{yr}}

\newcommand{\athena}{\texttt{Athena++}}
\newcommand{\grtrans}{\texttt{grtrans}}

\newcommand{\phn}{\phantom{0}}

\title[Jet Resolution]{The Effects of Resolution on Black Hole \\ Accretion Simulations of Jets}
\author[C.~J. White \& F. Chrystal]{
  Christopher J. White$^1$ and Fiona Chrystal$^2$ \\
  $^1$Kavli Institute for Theoretical Physics, University of California Santa Barbara, Kohn Hall, Santa Barbara, CA 93107, USA \\
  $^2$Department of Physics, University of California Santa Barbara, Broida Hall, Santa Barbara, CA 93106, USA
}
\date{Accepted ***. Received ***; in original form ***}
\pubyear{2020}

\begin{document}

\label{firstpage}
\pagerange{\pageref{firstpage}--\pageref{lastpage}}

\maketitle

\begin{abstract}
  We perform three general-relativistic magnetohydrodynamic simulations of black hole accretion designed to test how sensitive results are to grid resolution in the jet region. The cases differ only in numerics, modelling the same physical scenario of a radiatively inefficient, geometrically thick, magnetically arrested flow onto a rapidly spinning black hole. Properties inferred with the coarsest grid generally agree with those found with higher resolutions, including total jet power and its decomposition into different forms, velocity structure, nonaxisymmetric structure, and the appearance of resolved millimetre images. Some measures of variability and magnetization are sensitive to resolution. We conclude that most results obtained by limiting resolution near the jet for computational expediency should still be reliable, at least insofar as they would not be improved with a finer grid.
\end{abstract}


\section{Introduction}

Accreting black holes can launch relativistic jets that provide important probes of the physics near the event horizon. In particular, a combination of black hole spin and net magnetic flux aligned with the rotation axis can extract spin energy to power a jet \citep{Blandford1977}. This process can be important at scales ranging from stellar-mass black holes in the case of gamma-ray bursts to supermassive black holes in the case of active galactic nuclei.

Amenable conditions for jet launching involve large amounts of coherent magnetic flux pinned to the horizon by a geometrically thick accretion flow. This leads to a magnetically arrested disc (MAD), as predicted analytically \citep{BisnovatyiKogan1974} and seen in early numerical work \citep{Igumenshchev2003,Narayan2003}. The presence of an accretion flow made turbulent by the magnetorotational instability (MRI) and magnetic Rayleigh--Taylor instability forced further study of such systems to rely largely on magnetohydrodynamic (MHD) simulations. For example, \citet{Hawley2004} found a highly magnetized jet in general-relativistic (GR) MHD simulations of accretion onto a spinning black hole, \citet{Igumenshchev2008} illustrated the MAD--jet correspondence with pseudo-Newtonian calculations, and \citet{Tchekhovskoy2011} used GRMHD simulations to draw a clear connection between the MAD state around a spinning black hole and the extraction of spin energy to launch a jet. A large number of GRMHD calculations appear in the physical parameter survey of \citet{McKinney2012}. Further work has augmented these results with additional physics, such as optically thin cooling \citep{Avara2016}, radiative transfer \citep{McKinney2015,MoralesTeixeira2018}, and electron thermodynamics \citep{Ressler2017}.

With the ubiquity of global, three-dimensional, GRMHD simulations in this area, one hopes that the methods being used are converging to physically meaningful results. \Citet{White2019} began to address this specific concern with a resolution study of MAD discs. The focus was on physics near the midplane, and so layers of static mesh refinement (SMR) were used to successively refine a finite-volume grid in this region, measuring various aspects of the accretion flow to see whether they changed with resolution. In that study, most of the global properties converged at the highest resolution considered, though details of the turbulence and variability of synchrotron emission still showed signs of resolution dependence.

An important caveat to the \citeauthor{White2019} study is that it holds grid resolution fixed in the jet region. That work, like all of the aforementioned numerical investigations with the exception of \citet{Igumenshchev2003}, was performed on a spherical grid. Spherical coordinates (with logarithmic radial spacing) naturally concentrate resolution where needed for most accretion flows, and the ignorable azimuthal coordinate lends itself to angular momentum conservation about the respective axis, especially when the form of the conserved components of the stress-energy tensor are chosen well \citep{Gammie2003}. However, spherical grids tend to over-concentrate resolution near the polar axis. With $N_\theta$ cells equally spaced in the polar direction and $N_\phi$ cells in azimuth, the $\phi$-width of cells touching the axis scales as $N_\theta^{-1} N_\phi^{-1}$. Thus doubling the resolution in each dimension reduces the maximum stable timestep according to the Courant condition by a factor of $4$, as opposed to the factor of $2$ one achieves with Cartesian grids. In order to circumvent this problem, spherical grids are often adjusted to keep resolution low near the axis by using modified coordinates, irregular grid spacing, or mesh refinement. The latter is the case in \citet{White2019}.

Here we seek to complement the \citeauthor{White2019} study, focusing exclusively on the effects of varying resolution near the polar axis. We take the same physical setup, designed to reach a radiatively inefficient MAD state around a rapidly spinning black hole, and run it on three grids with different resolutions near the jet. The coarsest grid corresponds to the highest (midplane) resolution of \citet{White2019}, with the other grids successively refining the jet region while keeping the midplane grid fixed. Our three resolutions (detailed in Section~\ref{sec:numerical}) have $8$, $12$, and $24$ cells in polar angle $\theta$ at small radii across a fiducial half jet, from $\theta = 0$ to $\theta = 3 \pi / 16$, and they have $64$, $128$, and $256$ cells in azimuth, respectively. This numerical experiment enables us to assess how the resolution near the jet affects black hole accretion models.

Due to the aforementioned timestep scaling, running high spherical resolutions is generally prohibitively expensive. Most simulations in the literature use jet resolutions comparable to the coarsest considered here. For example, the A0.94BpN100 and A0.94BfN40 simulations of \citet{McKinney2012} have $6$ and $11$ cells in the polar direction inside the fiducial half jet, with $128$ and $256$ cells in azimuth. The highest resolution simulation (RADvHR) of \citet{MoralesTeixeira2018} has $3$ polar cells and $64$ azimuthal cells. \Citet{Ressler2017} have a very high resolution in polar angle -- $66$ cells in the half jet -- but they only have $64$ cells in azimuth.

In consideration of the cost, we do not run all three cases from early-time initial conditions, but rather evolve the system to steady state (at small radii) on the coarse grid and initialize the higher resolutions from that point. If we were modifying midplane resolution, one might worry that adjustments to the turbulence would necessitate adjustments to the steady-state flow on long, viscous timescales. However, here we are only interested in the behaviour of the jet, and to the extent it does not back-react on the bulk of the accreting material, we can expect adjustments, if any, of the jet to a new grid to occur on timescales comparable to the dynamical time of the inner accretion disc or the propagation time of the jet across the grid, both of which are short. These issues are discussed in more detail in Section~\ref{sec:discussion:relaxation}.

All of our simulations are performed with the finite-volume code \athena{} \citep{Stone2020}, using the full GRMHD equations \citep{White2016}. As we are employing a conservative code, we expect that even at low resolutions a jet will obtain the correct power. However, it remains possible that high resolutions are needed to resolve structure within the jet, or to reach the correct distribution of energy among the various forms such as electromagnetic or kinetic.

Throughout we refer to horizon-penetrating Kerr--Schild coordinates for a black hole of mass $M$ and spin parameter $a = 0.98$, either in spherical form $(t, r, \theta, \phi)$, or in Cartesian form $(t, x, y, z)$. The spatial coordinates are related via
\begin{subequations} \begin{align}
  x & = \sin\theta\  (r \cos\phi + a \sin\phi), \\
  y & = \sin\theta\  (r \sin\phi - a \cos\phi), \\
  z & = r \cos\theta.
\end{align} \end{subequations}
As the metric determinant is identically $-1$ for the Cartesian form, explicit appearances in formulas will always refer to the spherical determinant,
\begin{equation}
  \sqrt{-g} = \paren[\Big]{r^2 + a^2 \cos^2\!\theta} \sin\theta.
\end{equation}

Note we will refer to the $r$-direction as `radial;' this longitudinal coordinate along the jet is sometimes called `poloidal' in the jet literature. The $\theta$-direction we will refer to as `polar,' which in the jet is a lateral coordinate that the literature may call `radial.' When omitted, all quantities are expressed in units appropriately scaled to the black hole mass, for example lengths in units of $G M / c^2$. We use the Lorentz--Heaviside convention for the magnetic field, so for example magnetic pressure is $\mathbfit{B}^2 / 2$.

Section~\ref{sec:numerical} details our numerical setup. We present specific results on global accretion properties, jet power, jet structure, and the impact on millimetre observations in Section~\ref{sec:results}, with our findings summarized in Section~\ref{sec:discussion}.

\section{Numerical Methods}
\label{sec:numerical}

We begin with the highest-resolution piecewise parabolic method \citep[PPM]{Colella1984} simulation from \citet{White2019} at time $t = 10{,}000$, resetting the time coordinate to $t' = 0$. This simulation was initialized with a hydrostatic torus \citep{Fishbone1976} with inner edge at $r = 16.45$ and pressure maximum at $r = 34$, with adiabatic index $\Gamma = 13/9$. The rest-mass density $\rho$ is normalized to have an initial peak value of unity. The torus was seeded with a weak poloidal magnetic field that would accumulate a large net vertical flux over the coarse of the run.

The grid employed for that run consisted of four SMR levels, numbered $0$ through $3$, concentrating resolution toward the midplane. The effective resolution of the highest refinement level was $173$ cells per decade in radius, $256$ cells in polar angle, and $512$ cells in azimuthal angle.

\begin{figure}
  \centering
  \includegraphics{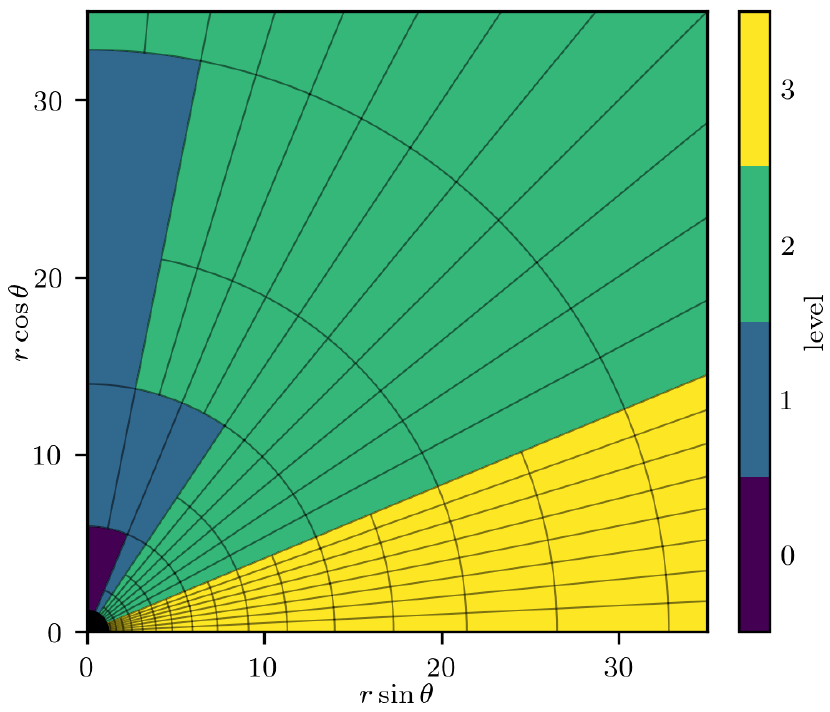}
  \caption{Poloidal slice of the grid used for the Level~0 simulation. The grid lines denote blocks of $16 \times 4$ cells. There are $2^l$ cells in radius interior to the horizon at level $l$. This is the same grid as shown in Figure~1 of \citet{White2019}. \label{fig:grid}}
\end{figure}

We continue with this simulation, denoting it Level~0 after its coarsest refinement level. We also run two additional simulations initialized at $t' = 0$ from this one but using different grids. The Level~1 simulation refines all of refinement level $0$ to level $1$, and the Level~2 simulation refines all of levels $0$ and $1$ to level $2$. Refinement level $l$ contains $64 \times 2^l$ cells in azimuth. The grid structure for the Level~0 simulation is shown in Figure~\ref{fig:grid}.

In all cases we continue with PPM reconstruction and the HLLE Riemann solver \citep{Einfeldt1988}. Relative to the lower order piecewise linear method also investigated in \citet{White2019}, PPM simulations are slightly closer to convergence for a fixed resolution.

\athena, like all ideal GRMHD codes, must impose nonzero density and gas pressure floors for numerical robustness. Given gas pressure $\pgas$, velocity $u^\mu$, and magnetic field $B^i$, the plasma magnetization parameters are calculated as
\begin{subequations} \begin{align}
  \sigma & = \frac{2 \pmag}{\rho}, \\
  \beta^{-1} & = \frac{\pmag}{\pgas},
\end{align} \end{subequations}
where we have $\pmag = b_\mu b^\mu / 2$, $b^t = u_i B^i$, and $b^i = (B^i + b^t u^i) / u^t$. We enforce the limits $\sigma \leq 100$ and $\beta^{-1} \leq 100$ throughout the simulation via floors on $\rho$ and $\pgas$ that are dependent on the magnetic field:
\begin{subequations} \begin{align}
  \rho & \geq \max\paren[\bigg]{10^{-8}, 10^{-4} r^{-3/2}, \frac{\pmag}{50}}, \\
  \pgas & \geq \max\paren[\bigg]{10^{-10}, 10^{-6} r^{-5/2}, \frac{\pmag}{100}}.
\end{align} \end{subequations}
When these floors are invoked, mass and/or gas pressure are added in the normal observer frame. We additionally limit the velocity to keep the normal-frame Lorentz factor from becoming too large:
\begin{equation}
  \gamma \equiv \alpha u^t \leq 50,
\end{equation}
with $\alpha \equiv (-g^{tt})^{-1/2}$ the lapse. The most prominent effect of these limits is the creation of extra mass at the stagnation surface at the base of the jet.

We run each setup to a time of at least $t' = 1930$. Per $1000$ simulation time units, these simulations cost $2700$, $9600$, and $36{,}000$ node-hours on $125$, $125$, and $133$ $68$-core Intel Xeon Phi 7250 (Knights Landing) nodes.

Throughout our analysis, we will find it useful to compare datasets generated by the different simulations. Given an unordered list of data elements (for example, the set of accretion rates seen in different snapshots), we will use standard Bayesian parameter estimation to report central values and uncertainties on the mean and standard deviation of the dataset. Samples will be treated as independent, and the calculation will be done employing the Jeffreys prior. The values will be reported as $X \pm S$, where $X$ is the mean of the posterior for the statistic (which is either the mean of the distribution or the standard deviation of the distribution), and $(X - S, X + S)$ is the median-centred $1$-sigma (i.e. approximately $68\%$) confidence interval on the posterior. Note that when the statistic is the standard deviation, the confidence interval is in theory asymmetric about the mean, but in fact this asymmetry does not appear with the number of significant figures we report.

\section{Results}
\label{sec:results}

\subsection{Accretion Rate and Magnetic Flux}

As even the Level~0 simulation is well resolved near the midplane, we do not expect to see statistical differences in accretion rates among the simulations. The region whose resolution is varying does not contribute to MRI turbulence. Similarly, the net amount of magnetic flux brought to the horizon should be controlled by the balance of accretion and magnetic buoyancy via the magnetic Rayleigh--Taylor instability near the midplane.

Define the accretion rate
\begin{equation}
  \mdot = -\oint\limits_{r=5} \rho u^r \sqrt{-g} \, \dd\theta \, \dd\phi
\end{equation}
and normalized horizon flux
\begin{equation} \label{eq:phi}
  \varphi = \frac{\sqrt{4 \pi}}{2 \sqrt{\mdot}} \oint\limits_{r=\rhor} \abs[\Big]{B^r} \sqrt{-g} \, \dd\theta \, \dd\phi.
\end{equation}
While the flux is measured at the horizon radius $\rhor = 1 + \sqrt{1 - a^2}$, accretion is measured at $r = 5$ in order to mitigate the effects of density floors artificially adding material inside the stagnation surface at high latitudes. The typical MAD state saturates near $\varphi \approx 47$ \citep{Tchekhovskoy2011} or $56$ \citep{White2019}.

\begin{figure}
  \centering
  \includegraphics{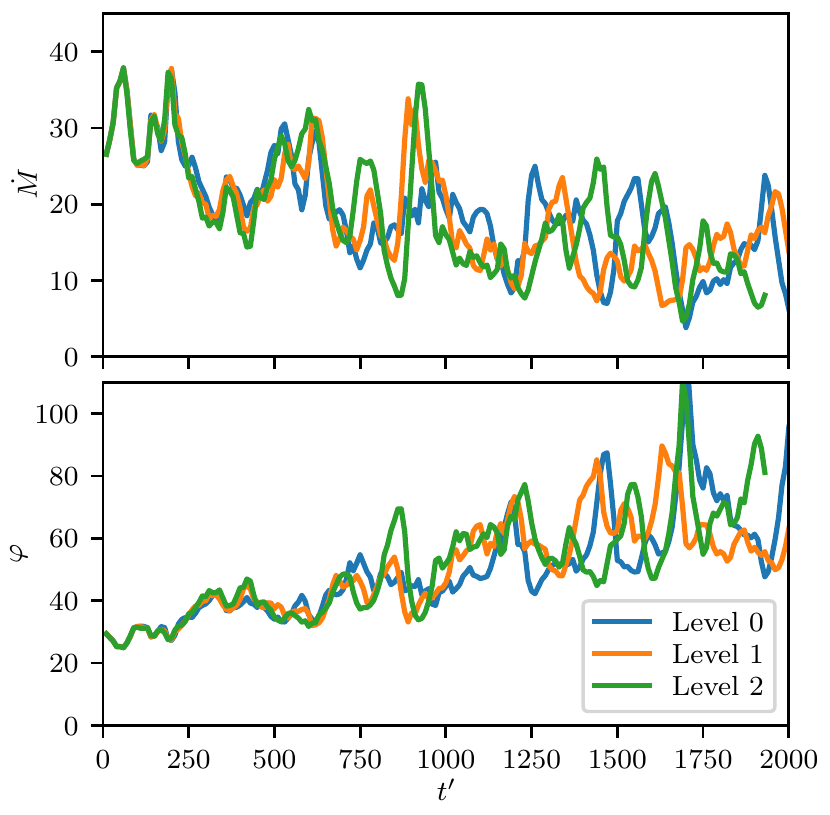}
  \caption{Accretion rate and normalized horizon magnetic flux as functions of time. The Level~1 and Level~2 simulations are initialized at $t' = 0$, $10{,}000$ time units after the Level~0 simulation is initialized. The curves largely trace one another in a statistical sense, though there is more variability in $\mdot$ at higher resolutions. \label{fig:accretion}}
\end{figure}

As can be seen in Figure~\ref{fig:accretion}, there are no large systematic differences in behaviour among the three simulations with regard to either $\mdot$ or $\varphi$. Both vary about approximately the same steady-state values, with flux levels in agreement with the literature.

We quantify the statistical properties of each of these curves using parameter estimation as described in Section~\ref{sec:numerical}. Here we use the $144$ samples for $500 \leq t' \leq 1930$, by the beginning of which point the curves have begun to diverge from their initial conditions (see Section~\ref{sec:discussion:relaxation} for justification). The central values and uncertainties for the means and standard deviations of the datasets are reported in Table~\ref{tab:accretion}.

\begin{table}
  \centering
  \caption{Global accretion statistical properties. \label{tab:accretion}}
  \begin{tabular}{cccc}
    \toprule
    Quantity & Simulation & Mean & Std.\ Dev. \\
    \midrule
    \multirow{3}{*}{$\mdot$} & Level~0 & $17.02 \pm 0.46$ & $5.34 \pm 0.33$ \\
    & Level~1 & $16.40 \pm 0.51$ & $6.12 \pm 0.36$ \\
    & Level~2 & $16.73 \pm 0.59$ & $7.12 \pm 0.42$ \\
    \midrule
    \multirow{3}{*}{$\varphi$} & Level~0 & $54.4 \pm 1.3$ & $15.14 \pm 0.90$ \\
    & Level~1 & $55.3 \pm 1.1$ & $13.38 \pm 0.79$ \\
    & Level~2 & $56.5 \pm 1.3$ & $15.61 \pm 0.92$ \\
    \bottomrule
  \end{tabular}
\end{table}

The mean values in the third column are all in agreement:\ the magnitude of the difference between any two central values is similar to or greater than the square root of the sum of the squares of the uncertainties. However, this is not the case with the standard deviations in the fourth column. While there is no clear trend with resolution for $\varphi$, variability in $\mdot$ increases systematically with resolution. There is a $3.0$-sigma difference between the standard deviations in $\mdot$ for Levels~0 and~2. Resolving the base of the jet results in larger amplitudes of time variability in accretion rate near the horizon.

\subsection{Jet Power}
\label{sec:results:jet_power}

We now turn to properties directly measured from the jet itself. Following \citet{McKinney2012}, we decompose the stress-energy tensor into electromagnetic, kinetic, enthalpy, and rest mass terms:
\begin{subequations} \begin{align}
  \tensor{T}{^\mu_\nu} & = \tensor{(\tem)}{^\mu_\nu} + \tensor{(\tkin)}{^\mu_\nu} + \tensor{(\ten)}{^\mu_\nu} + \tensor{(\tmass)}{^\mu_\nu}, \\
  \tensor{(\tem)}{^\mu_\nu} & = 2 \pmag u^\mu u_\nu + \pmag \delta^\mu_\nu - b^\mu b_\nu, \\
  \tensor{(\tkin)}{^\mu_\nu} & = \rho u^\mu (u_\nu + \delta^t_\nu), \\
  \tensor{(\ten)}{^\mu_\nu} & = (\ugas + \pgas) u^\mu u_\nu + \pgas \delta^\mu_\nu, \\
  \tensor{(\tmass)}{^\mu_\nu} & = -\rho u^\mu \delta^t_\nu.
\end{align} \end{subequations}
Here $\ugas = \pgas / (\Gamma - 1)$. Define the outward power in the jet at a particular time and radius as
\begin{equation}
  \pjet = -\int\limits_\mathrm{jet} \tensor{T}{^r_t} \sqrt{-g} \, \dd\theta \, \dd\phi,
\end{equation}
applicable to either the total stress-energy or any of its constituents. The integration is restricted to be over the jet region, defined to be where $\sigma > 1$.

\begin{figure}
  \centering
  \includegraphics{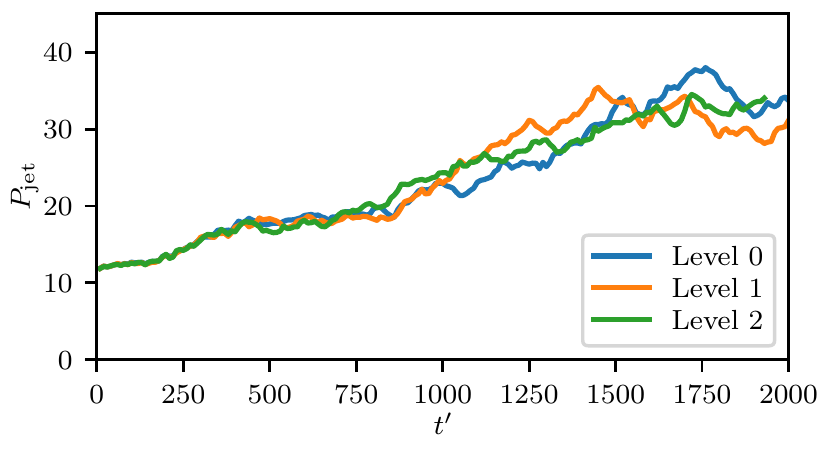}
  \caption{Total outward jet power for the three simulations. The three curves follow each other closely. The long-term trend reflects the same secular trend seen in $\varphi$ in Figure~\ref{fig:accretion}. \label{fig:jet_power_time}}
\end{figure}

The run of total jet power with time for all three simulations is shown in Figure~\ref{fig:jet_power_time}. Here we fix the radius to $r = 50$. There is a secular increase in power over the time shown, though we have no reason to believe this is anything other than a stochastic fluctuation. There are no striking differences among the three curves.

We can quantify the time-variability in the jet power as follows. Take all data for $500 \leq t' \leq 1930$. Divide the data into all $130$ possible (overlapping) chunks of $15$ contiguous samples (the sample spacing being $\Delta t = 10$). Within each chunk, calculate the standard deviation of the $15$ data values, divided by the mean of the same. Let $\sigma_P$ be the mean of these $130$ values, capturing the variability in the jet power. We find $\sigma_P$ to be $0.036$, $0.037$, and $0.031$ for Level~0, Level~1, and Level~2, respectively. That is, short-time variability shows no particular trend with resolution. This is seen even with other chunk sizes.

\begin{figure*}
  \includegraphics{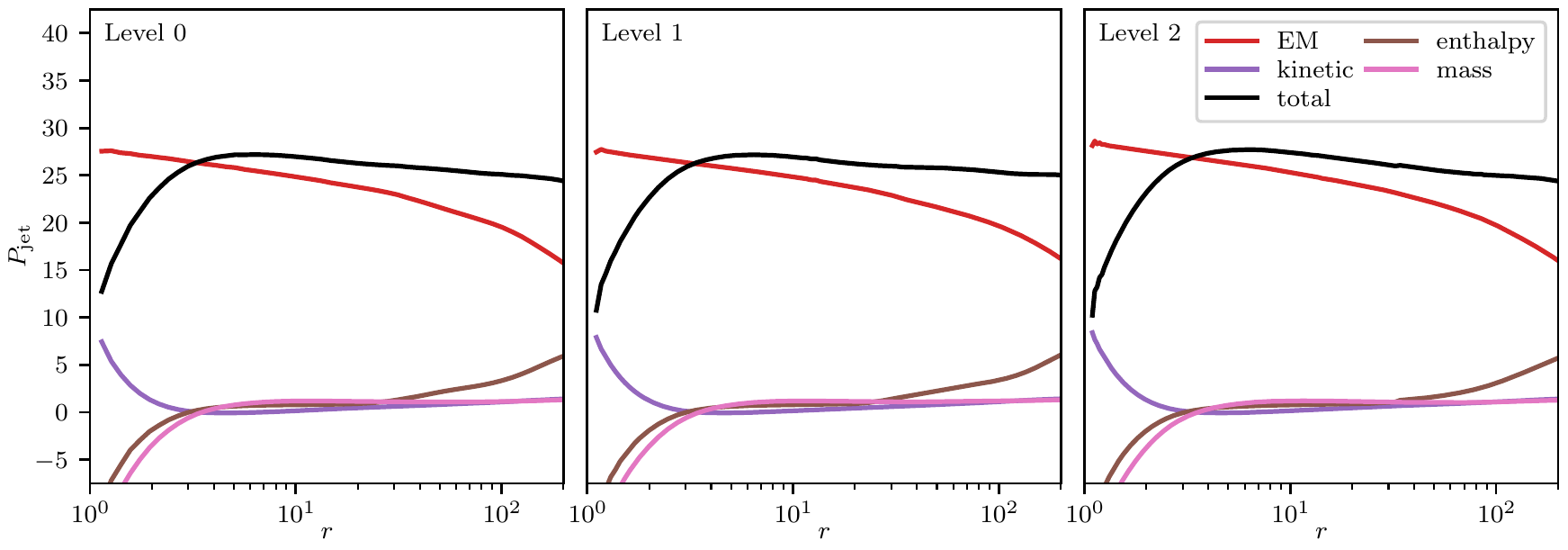}
  \caption{Radial profiles of jet power for the three simulations, averaged over $500 \leq t' \leq 1930$. There is very close agreement not only in total power but also in the decomposition of that power. \label{fig:jet_power_radius}}
\end{figure*}

Instead of time series of radial slices, we can consider time-averaged radial profiles of $\pjet$. Figure~\ref{fig:jet_power_radius} shows these profiles for all three cases, showing the decompositions as well as the total power. The profiles are averaged over the time $500 \leq t' \leq 1930$. The three panels are remarkably similar, with all three jets electromagnetically dominated, though with electromagnetic energy converting to gas internal energy at large radii. Note that even in steady state neither the total power nor the rest mass flux need be constant with radius, given that we are only capturing the fluxes in the $\sigma > 1$ region.

While much of the mass in the jet on both sides of the stagnation surface may be created as part of enforcing $\sigma \leq 100$, this does not have a large effect on the total emitted power. The floors only affect regions that are very electromagnetically dominated, where most of the power is carried by the Poynting flux. Lowering the $\sigma$ ceiling from $100$ to $10$, for example, might change the nature of the jet, but increasing it to $1000$ could not have too large a consequence, since we are already near to the force-free limit.

\begin{figure}
  \centering
  \includegraphics{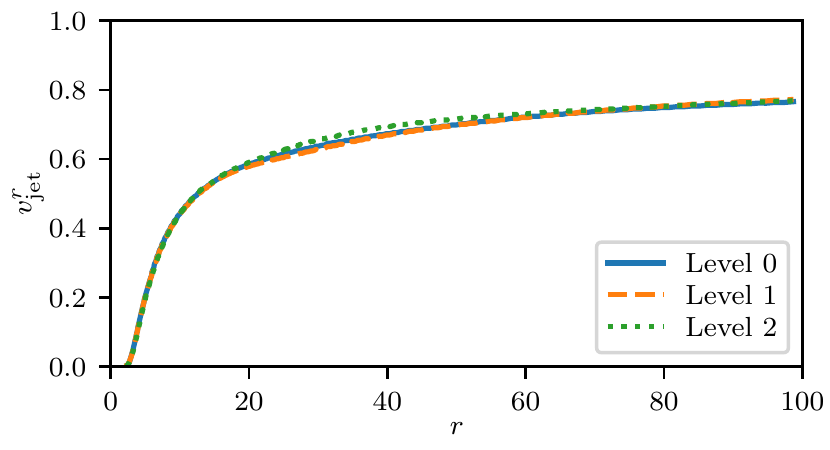}
  \caption{Radial profiles of jet velocity for the three simulations, averaged over $500 \leq t' \leq 1930$. The close agreement here is expected given the agreement in energy fluxes shown in Figure~\ref{fig:jet_power_radius}. \label{fig:velocity}}
\end{figure}

The agreement among the simulations regarding the longitudinal structure of the jet is reflected in the velocity. Define the time-averaged $r$-velocity
\begin{equation}
  \vjet = \frac{\int_{500}^{1930} \! \int_\mathrm{jet} \paren[\Big]{u^r / u^t} \sqrt{-g} \, \dd\theta \, \dd\phi \, \dd t'}{\int_{500}^{1930} \! \int_\mathrm{jet} \sqrt{-g} \, \dd\theta \, \dd\phi \, \dd t'}.
\end{equation}
Note this is the actual velocity of the plasma in the simulation, not an asymptotic velocity obtained by assuming all energy will eventually become kinetic at large radii as is sometimes reported in the literature. Figure~\ref{fig:velocity} shows the run of $\vjet$ with radius for the three simulations. The profiles are essentially indistinguishable. This also holds (though with slightly different profiles) if we weight the velocity by $\rho$ or $\sigma$ when averaging.

Unlike with total power, we expect the velocity of the jet to be affected by numerical floors to a large degree. Our $\sigma$ ceiling effectively adds mass in the normal observer frame, while other choices made in the literature include the fluid frame and the drift frame defined in \citet{Ressler2017}. The prescriptions adopted by other codes may well yield different velocity profiles, but we have no reason to believe they would be more sensitive to resolution than ours. If the floors were needed only at isolated points in spacetime, then one might expect that as resolution is increased (and the timestep is correspondingly decreased) the spacetime volume of cells directly modified by flooring routines would decrease, and so the impact of the floors would change in magnitude. The fact that we see the same profiles at different resolutions indicates this is not the case;\ the floors are largely applied systematically over a region, and the aggregate effect, while sensitive to the choice of flooring algorithm, does not depend on resolution.

\subsection{Lateral Structure}

The structure of the jet in the lateral direction ($\theta$) is determined by the launching conditions near the black hole and interaction with matter as it propagates. As the former is typically calculated in regions with very coarse spatial resolution, one might worry that jets are imprinted with inaccurate conditions upon launching. Here we examine three averaged quantities as functions of lateral position:\ the normal-frame Lorentz factor $\gamma$, plasma $\beta^{-1}$, and plasma $\sigma$.

\begin{figure*}
  \includegraphics{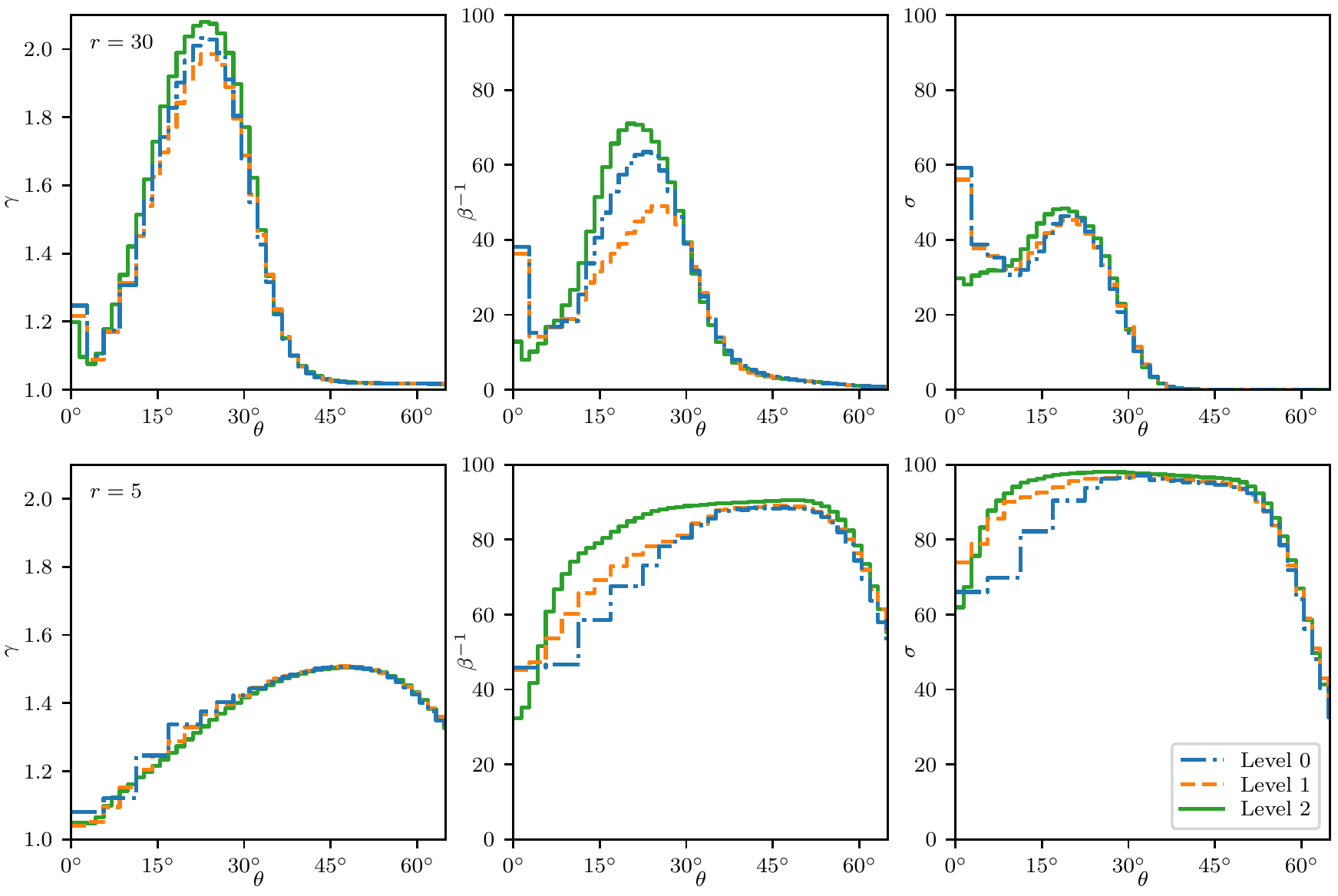}
  \caption{Lateral profiles of Lorentz factor and magnetization for the three simulations, averaged over $500 \leq t' \leq 1930$ and sampled at two different radii. The horizontal segments correspond to annuli that are a single cell thick in the respective simulations. The velocity structure is relatively independent of resolution, but the magnetization levels tend to be higher in the outer parts of the jet and lower in the core at higher resolution. \label{fig:lateral}}
\end{figure*}

Each quantity is averaged in time over $500 \leq t' \leq 1930$ and in azimuth, and the results from the northern and southern jets are averaged. Figure~\ref{fig:lateral} shows the profiles at two different radii, out to an angle of $65^\circ$ away from the axis. At $r = 30$, the Level~0 and Level~1 simulations cover this region with refinement levels $1$ and $2$. At $r = 5$, the Level~0 simulation uses refinement levels $0$ through $2$, while Level~1 uses levels $1$ and $2$. In both cases, the Level~2 simulation uses refinement level $2$ everywhere in the jet region. The discrete grid structure is reflected in the step-function nature of the curves.

The velocity structure of the jet is little affected by resolution. In all cases, the jet develops a fast sheath around a slow core as it propagates to large radii. While the sheath might be slightly faster at highest resolution compared to the other two cases, it is slowest at the middle resolution;\ there is no clear trend. Note that even in the fastest regions we have $\gamma \lesssim 2$, far from the numerical ceiling of $50$.

On the other hand, there is less agreement in terms of magnetization. Both $\beta^{-1}$ and $\sigma$ are systematically larger outside the core at higher resolutions. It is likely that low resolutions suffer from large numerical reconnection, artificially draining the jet of its magnetic energy as it propagates outward. While a simulation that sacrifices resolution near the jet may get the correct total power, it may well be inaccurate in terms of what forms that power takes and how energy and other properties vary across a cross section. Moreover, the apparent agreement in $\sigma$ between resolutions at small radii beyond approximately $20^\circ$ away from the axis is due to the curves approaching the limiting value of $100$ imposed numerically. A less restrictive cut-off would likely lead to greater discrepancy between resolutions in this region.

\subsection{Spiral Structure}

At any one snapshot in time, the jets in our simulations are not perfectly symmetric about the polar axis, and this asymmetry smoothly rotates around the axis as one moves in the longitudinal ($r$) direction.

This spiral structure can be seen in the average position of the jet as a function of $t$ and $r$. At each time and radius and for each of the two jets, define the jet region to be all cells with $u^r > 0$, $\sigma > 1$, and $\theta \leq \pi / 4$ or $\theta \geq 3 \pi / 4$ as appropriate. The $\sigma$-weighted moment of the Cartesian Kerr--Schild coordinate $x$ is
\begin{equation}
  \bar{x} = \frac{\int_\mathrm{jet} x \sigma \sqrt{-g} \, \dd\theta \, \dd\phi}{\int_\mathrm{jet} \sigma \sqrt{-g} \, \dd\theta \, \dd\phi},
\end{equation}
and the $y$-moment is defined similarly.

\begin{figure}
  \centering
  \includegraphics{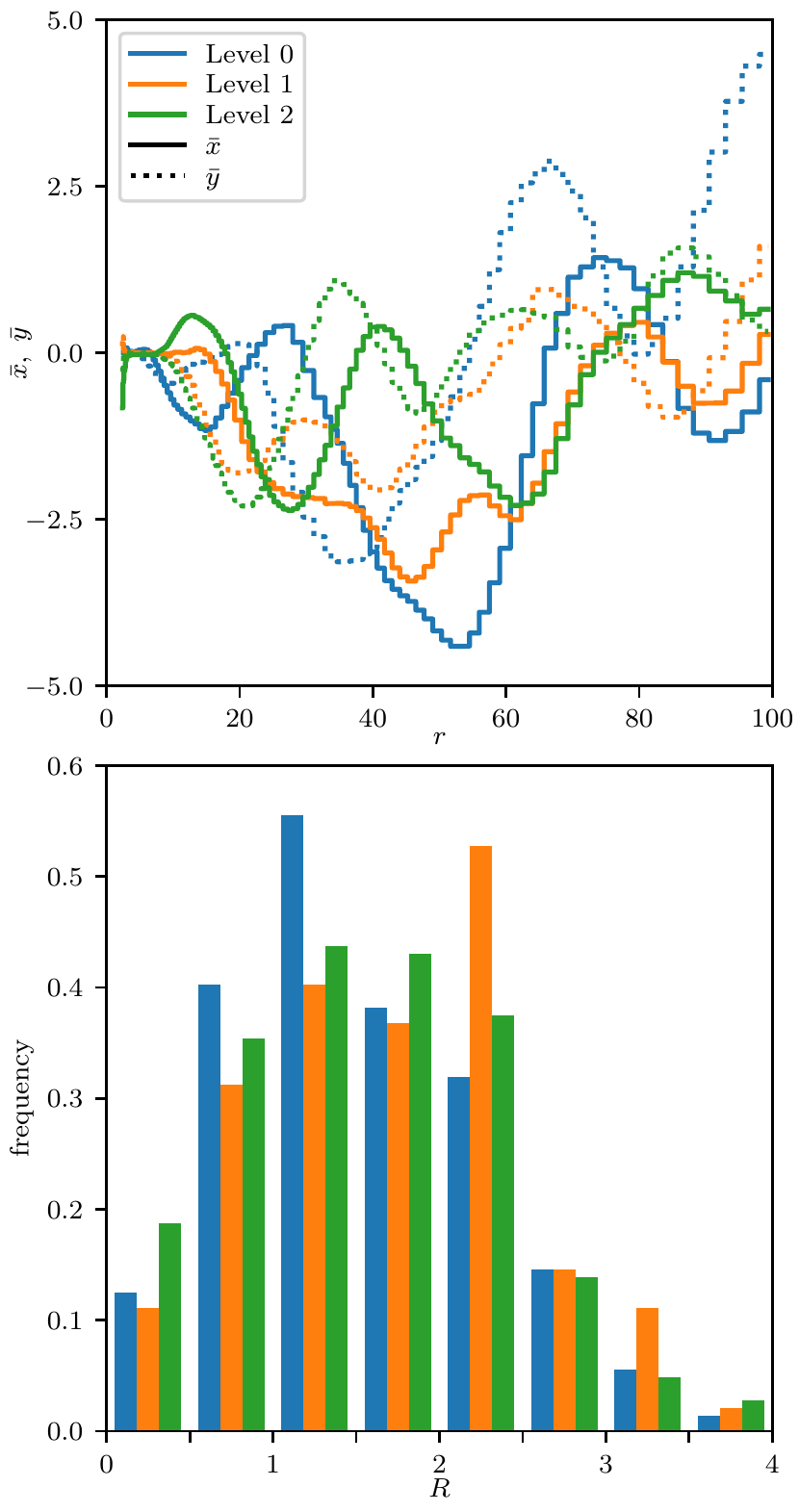}
  \caption{Top:\ jet centre coordinates $\bar{x}$ and $\bar{y}$ along the jet at $t' = 1000$ in all three simulations. The steps indicate the radial extent of the cells along the axis in the corresponding grid. The relative phases match what would be expected for structure imprinted at the base of the jet by a feature moving in the prograde direction. Bottom:\ histogram of lateral displacements $R$ at $r = 30$, using data from all snapshots $500 \leq t' \leq 1930$ and from both jets. \label{fig:spiral}}
\end{figure}

The top panel of Figure~\ref{fig:spiral} shows the runs of $\bar{x}$ and $\bar{y}$ with $r$ in the northern jet of all three simulations at $t' = 1000$. There is an oscillatory behaviour in all cases. Moreover, the relative phases of $\bar{x}$ and $\bar{y}$ indicate that the $\sigma$-weighted jet centre spirals about the axis in the prograde direction (anticlockwise as viewed from the north) as one moves toward the midplane. The southern jet (not shown) also displays spiral structure with a prograde sense of rotation moving toward the midplane. These are consistent with structure first being imprinted on the jet near the midplane by a feature moving with the accretion flow, then being advected away with the flow of the jet.

The displacements of the jet centres from the axis at $r = 30$ are shown in the lower panel of Figure~\ref{fig:spiral}. Here we simply measure displacement as
\begin{equation}
  R = \sqrt{\bar{x}^2 + \bar{y}^2}.
\end{equation}
The histogram uses all $144$ snapshots covering $500 \leq t' \leq 1930$, and it makes use of both jets. As with the global accretion properties, we summarize the statistics of these distributions with a central value and $1$-sigma confidence interval for the mean and standard deviation. These values are given in Table~\ref{tab:spiral}. There is no strong trend with resolution;\ even the coarse grid is able to capture spiral structure in the jet. While the Level~1 simulation shows a statistically significantly higher mean displacement than the other two, the fact that this occurs at an intermediate resolution suggests the difference may simply be due to stochastic variations on long timescales, rather than being a numerical effect.

\begin{table}
  \centering
  \caption{Statistical properties of lateral displacement $R$ at $r = 30$. \label{tab:spiral}}
  \begin{tabular}{ccc}
    \toprule
    Simulation & Mean & Std.\ Dev. \\
    \midrule
    Level~0 & $1.527 \pm 0.043$ & $0.739 \pm 0.031$ \\
    Level~1 & $1.708 \pm 0.047$ & $0.792 \pm 0.033$ \\
    Level~2 & $1.553 \pm 0.045$ & $0.772 \pm 0.032$ \\
    \bottomrule
  \end{tabular}
\end{table}

\subsection{Ray Tracing}

Given the current importance of resolved millimetre observations in comparing supermassive black hole accretion flows to models, we investigate whether varying resolution in the jet impacts the predicted images.

We use the general-relativistic ray tracing code \grtrans{} \citep{Dexter2009,Dexter2016} to create images from our simulation snapshots at observer frequencies of $230$ and $43\ \ghz$. The camera is placed at radial coordinate $r = 100$ with a field of view of either $20$ (at $230\ \ghz$) or $40$ gravitational radii, covering the image plane with $128^2$ rays. We use the same black hole parameters as in \citet{EHT2019f} appropriate for M87:\ a mass of $6.5 \times 10^9\ \msun$ and a distance of $16.8\ \mpc$. In order to obtain electron temperatures from the ideal, single-fluid simulations, we use the proton-to-electron temperature ratio prescription from \citet{Moscibrodzka2016},
\begin{equation}
  \frac{\tp}{\te} = \frac{\rhigh \beta^2 + \rlow}{1 + \beta^2}
\end{equation}
with $\rhigh = 10$ and $\rlow = 1$, fixing all emission and absorption to be thermal synchrotron from the electrons.

The single remaining free parameter -- the density scale of the simulation -- is adjusted independently for the three resolutions in order for the average flux from $144$ snapshots over $500 \leq t' \leq 1930$ to match the observed value of $0.98\ \jy$ at $230\ \ghz$ \citep{Doeleman2012}. This parameter takes on very similar values for the three cases. Combined with the code unit accretion rates in Table~\ref{tab:accretion}, we find physical accretion rates of $6.6$, $6.6$, and $6.5 \times 10^{-5}\ \msun/\yr$ for Levels 0 through 2, respectively. These correspond to $4.6$, $4.6$, and $4.5 \times 10^{-7}$ of the Eddington rate, defined here in terms of the proton mass and Thompson cross section as $\mdotedd = 10 \cdot 4 \pi G M \mproton / c \sigmat$.

When performing such ray tracing, it is common to exclude regions of the simulation with $\sigma > 1$, treating them as vacuum. This is done to avoid contaminating results with the presumably artificially mass-loaded, highly magnetized jet. Here, though, we are interested in whether the simulation makes the same predictions at higher resolutions, whether or not those predictions are dependent on this numerical caveat. Thus we take the simulation data at face value, rather than masking what would amount to most of the jet.

\begin{figure*}
  \includegraphics{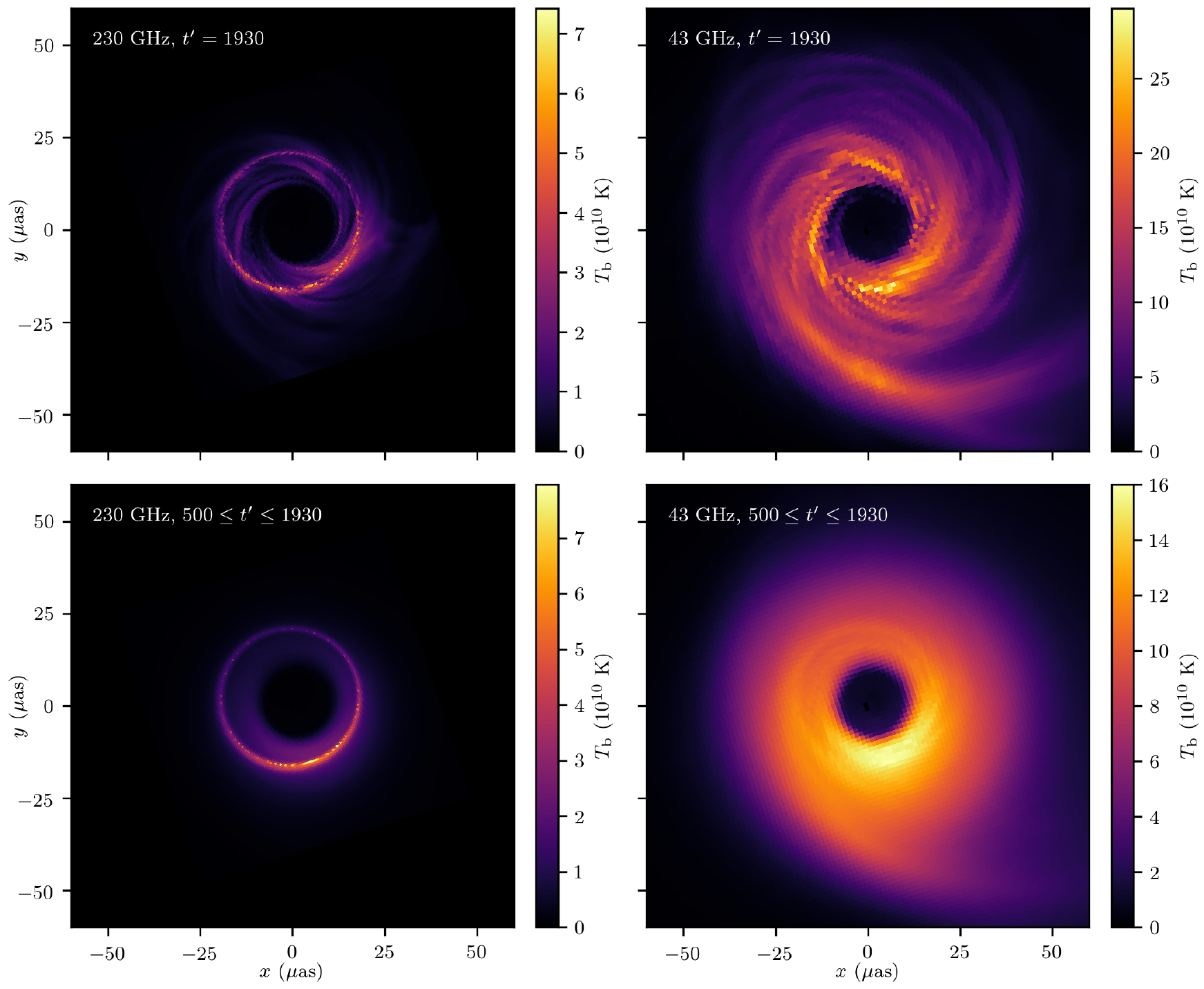}
  \caption{Ray tracing brightness temperature images from the Level~2 simulation. The black hole and camera parameters are chosen to match M87. The top row shows snapshots at the end of the simulation, and the bottom row shows averages over the entire time after the higher resolution simulations were initialized. The left panels show $230\ \ghz$ emission (used for calibration), while the right panels show $43\ \ghz$ images (which emphasize the jet). \label{fig:images_face}}
\end{figure*}

Figure~\ref{fig:images_face} shows the resulting images from the Level~2 simulation at both frequencies, both from the last snapshot and averaged over all $144$ snapshots. The camera is positioned and rotated such that the southern black hole spin axis matches the observed large-scale radio jet in M87, pointed toward the viewer, $17^\circ$ off the line of sight and $18^\circ$ north of west \citep{Mertens2016,Walker2018}.

At $43\ \ghz$ much of the emission is coming from the jet and its boundary layer. With a viewing angle mostly aligned with the jet, we can see transient spiral structures in the image (upper right panel). These are far enough outside the photon ring that they must be the direct result of spatial variations in the fluid rather than highly curved photon geodesics. As expected, these features are missing from the time-averaged image (lower right panel). Moreover, the averaged images from the other resolutions (not shown) look nearly indistinguishable.

\begin{figure}
  \centering
  \includegraphics{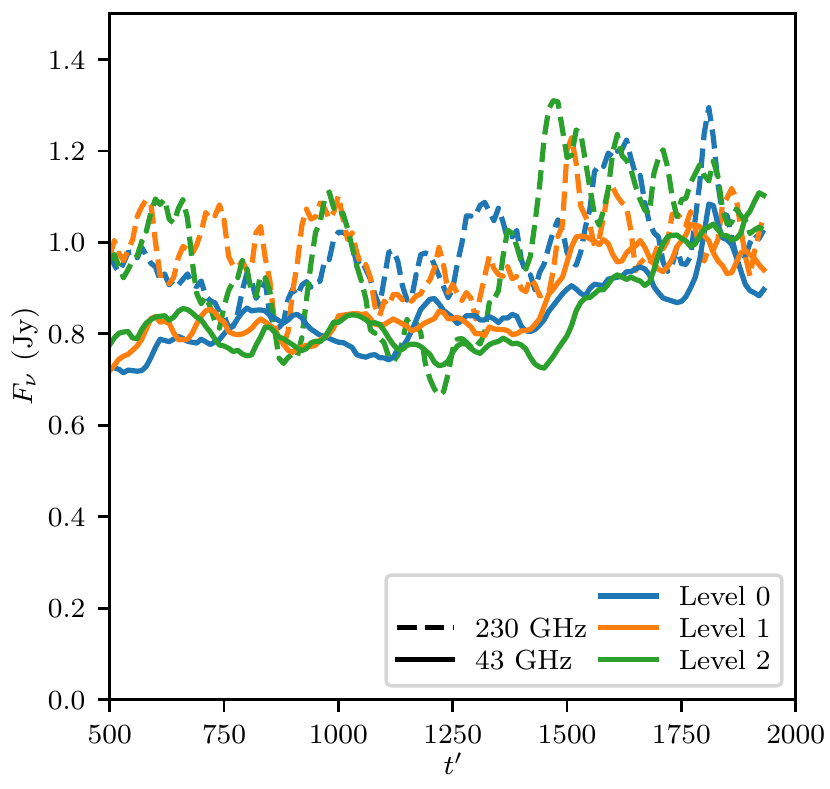}
  \caption{Ray tracing light curves for all three simulations using parameters appropriate for M87. The mean fluxes at $43\ \ghz$ are in agreement (the curves are calibrated to have the same $230\ \ghz$ flux), but at both frequencies there is more variability at higher resolution. \label{fig:light_curves_face}}
\end{figure}

Integrating the intensity in each frame yields light curves, shown in Figure~\ref{fig:light_curves_face}. At both frequencies, the light curves from the three simulations appear to track one another reasonably well. Table~\ref{tab:light_curves_face} quantifies these properties via parameter estimation for the mean and standard deviation. Here we can see variability systematically increasing with resolution. The Level~2 standard deviation differs from the Level~0 value by $5.4$ sigma at $230\ \ghz$ and by $3.0$ sigma at $43\ \ghz$.

\begin{table}
  \centering
  \caption{Statistical properties of light curves modelling M87 as viewed from Earth. \label{tab:light_curves_face}}
  \begin{tabular}{cccc}
    \toprule
    Frequency & Simulation & Mean ($\jy$) & Std.\ Dev. ($\jy$) \\
    \midrule
    \multirow{3}{*}{$230\ \ghz$} & Level~0 & $0.9802 \pm 0.0081$ & $0.0974 \pm 0.0058$ \\
    & Level~1 & $0.9800 \pm 0.0070$ & $0.0836 \pm 0.0049$ \\
    & Level~2 & $0.980\phn \pm 0.013\phn$ & $0.1566 \pm 0.0093$ \\
    \midrule
    \multirow{3}{*}{$43\ \ghz$} & Level~0 & $0.8419 \pm 0.0062$ & $0.0744 \pm 0.0044$ \\
    & Level~1 & $0.8652 \pm 0.0070$ & $0.0841 \pm 0.0050$ \\
    & Level~2 & $0.8434 \pm 0.0080$ & $0.0963 \pm 0.0057$ \\
    \bottomrule
  \end{tabular}
\end{table}

\begin{figure*}
  \includegraphics{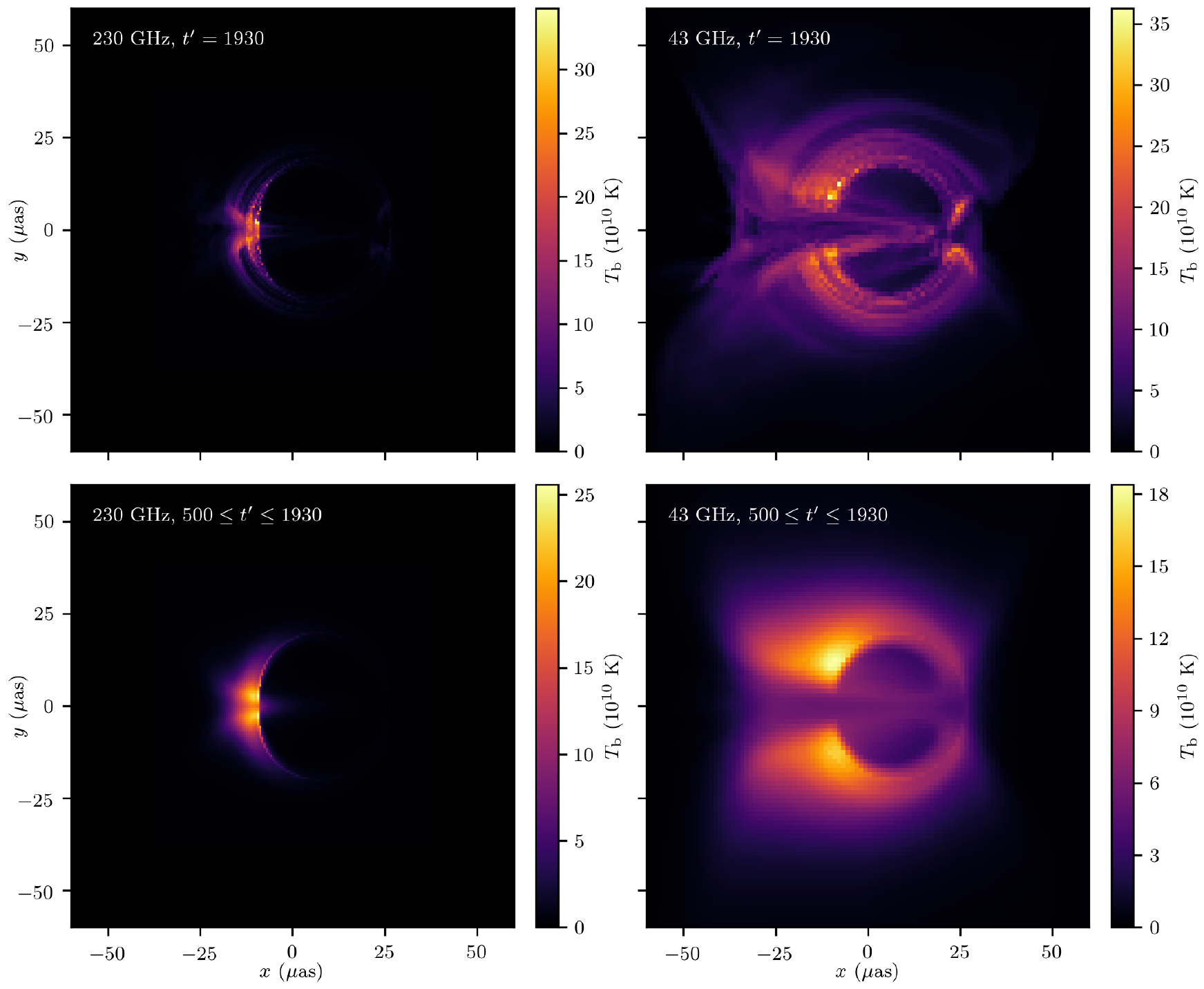}
  \caption{Ray tracing brightness temperature images from the Level~2 simulation. Relative to Figure~\ref{fig:images_face}, the camera is moved to view the accretion flow edge-on. The top and bottom rows show single snapshots and time averages, respectively, while the left and right columns show $230$ and $43\ \ghz$ images, respectively. Most of the emission at $43\ \ghz$ comes from high latitudes near the base of the jet. \label{fig:images_edge}}
\end{figure*}

It is possible that resolution in the jet region affects mock images at other viewing angles. In order to investigate this, we perform the same ray tracing on all snapshots at all resolutions with the camera moved to the midplane. All other parameters are kept fixed. Figure~\ref{fig:images_edge} shows the corresponding ray tracing images.

As before, there is an extended region of emission from in and near the jet at the lower frequency. The upper right panel again shows structure that must arise from spatial variations near where we are changing resolution. Still, such structures are transient, and the time-averaged images for the other simulations (not shown) are essentially the same as the lower right panel.

\begin{figure}
  \centering
  \includegraphics{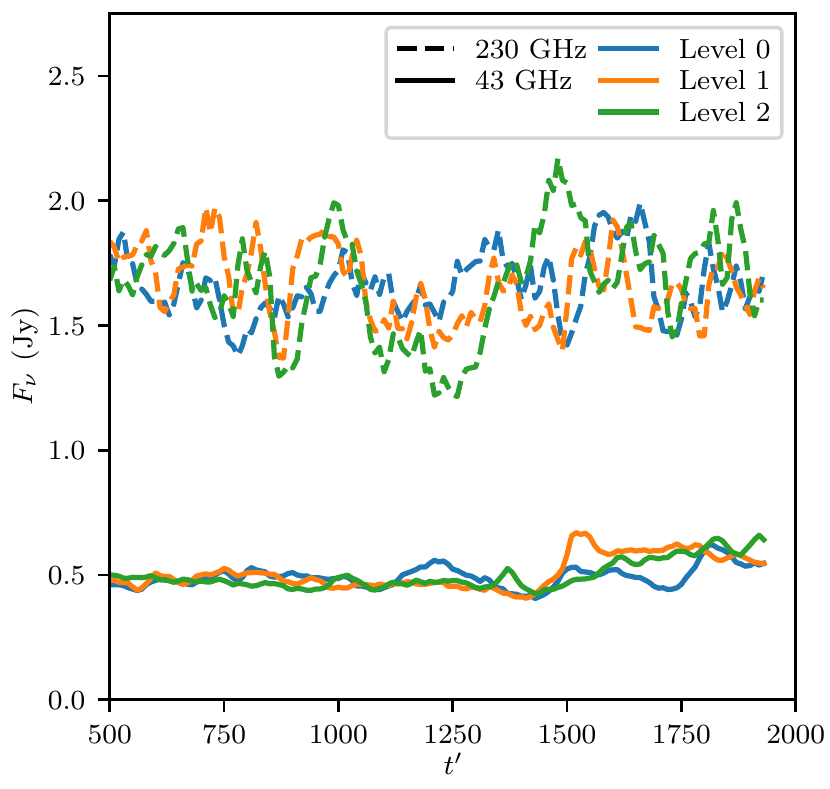}
  \caption{Ray tracing light curves for all three simulations using parameters appropriate for M87 but with a line of sight in the midplane. As with the nearly face-on viewing angle, the curves roughly agree at both frequencies, with variability increasing with resolution. Here there is a much larger offset between the two sets of curves. \label{fig:light_curves_edge}}
\end{figure}

Again, we generate the light curves, shown in Figure~\ref{fig:light_curves_edge}, and calculate some statistical properties, listed in Table~\ref{tab:light_curves_edge}. The mean fluxes show no strong trend with resolution, but once again the variability does appear to increase with resolution. At $230\ \ghz$, the standard deviations are nearly $70\%$ larger at Level~2 relative to Level~0, a $5.8$-sigma difference. At $43\ \ghz$, the Level~1 and Level~2 standard deviations are close ($1.7$ sigma), while they are significantly higher than what is found for Level~0 ($4.9$ and $3.4$ sigma).

\begin{table}
  \centering
  \caption{Statistical properties of light curves modelling M87 as viewed from the midplane. \label{tab:light_curves_edge}}
  \begin{tabular}{cccc}
    \toprule
    Frequency & Simulation & Mean ($\jy$) & Std.\ Dev. ($\jy$) \\
    \midrule
    \multirow{3}{*}{$230\ \ghz$} & Level~0 & $1.659 \pm 0.011$ & $0.1278 \pm 0.0076$ \\
    & Level~1 & $1.657 \pm 0.012$ & $0.1452 \pm 0.0086$ \\
    & Level~2 & $1.665 \pm 0.018$ & $0.214\phn \pm 0.013\phn$ \\
    \midrule
    \multirow{3}{*}{$43\ \ghz$} & Level~0 & $0.4923 \pm 0.0036$ & $0.0429 \pm 0.0025$ \\
    & Level~1 & $0.5085 \pm 0.0055$ & $0.0659 \pm 0.0039$ \\
    & Level~2 & $0.4988 \pm 0.0048$ & $0.0571 \pm 0.0034$ \\
    \bottomrule
  \end{tabular}
\end{table}

Whether viewed close to face-on, as is the case with M87 itself, or edge-on, resolution effects near the base of the jet do not lead to appreciable differences in time-averaged fluxes. Higher resolution does, however, allow for larger amplitudes of time variability, at least for timescales between $10$ and $1000$ gravitational times.

\section{Discussion}
\label{sec:discussion}

\subsection{Relaxation of the Simulations to New Grids}
\label{sec:discussion:relaxation}

In cases where we find that high resolution has little or no advantage, one might worry that the similarities among the different runs are merely reflective of not having enough time to diverge in character. Our comparisons begin at $t' = 500$, and the question is whether this is a long enough settling time for the Level~1 and Level~2 simulations to adjust to their new grids. In addition to the fact that we do find some differences over the timespan used, there are a priori and a posteriori arguments justifying our choice.

While it is true that GRMHD disc simulations are often run for at least $5000$ gravitational times, and sometimes for more than double that, before they are declared to have reached steady state, this corresponds to viscous relaxation timescales in the disc. As the viscous time (and even the orbital time) scales with radius as $r^{3/2}$, the radius inside of which steady state is achieved scales as $t^{2/3}$. Unlike in \citet{White2019}, however, we are not modifying the grid near the disc, nor are we directly interested in MRI-driven transport of angular momentum. Insofar as we only care about the jet, which should be determined by conditions at small radii where it is launched, $500$ gravitational times should be sufficient for relaxation. This corresponds to $26$ geodesic orbital periods at the innermost stable circular orbit.

One could argue that true steady state requires any back-reactions from the jet on the disc to be taken into account. That is, if changes to the grid alter the jet, which in turn alters the disc at large radii, then one must wait for these influences to viscously return to the jet (multiple times in fact) to find the equilibrium. In aligned systems such as the ones we consider here, where the jet never points toward the infalling material, such feedback is expected to be small. Indeed, the vast majority of simulations of this type only approach steady state out to a few tens of gravitational radii at most;\ any effects of the jet on the disc at larger radii are rarely modelled. Moreover, we can see from Figure~\ref{fig:jet_power_time} that the effects of changing resolution do not include large alterations to the strength of the jet. That is, what little back-reaction there is has had $10{,}000$ gravitational times before $t' = 0$ to establish an equilibrium, and that equilibrium is not expected to change after $t' = 0$.

\begin{figure}
  \centering
  \includegraphics{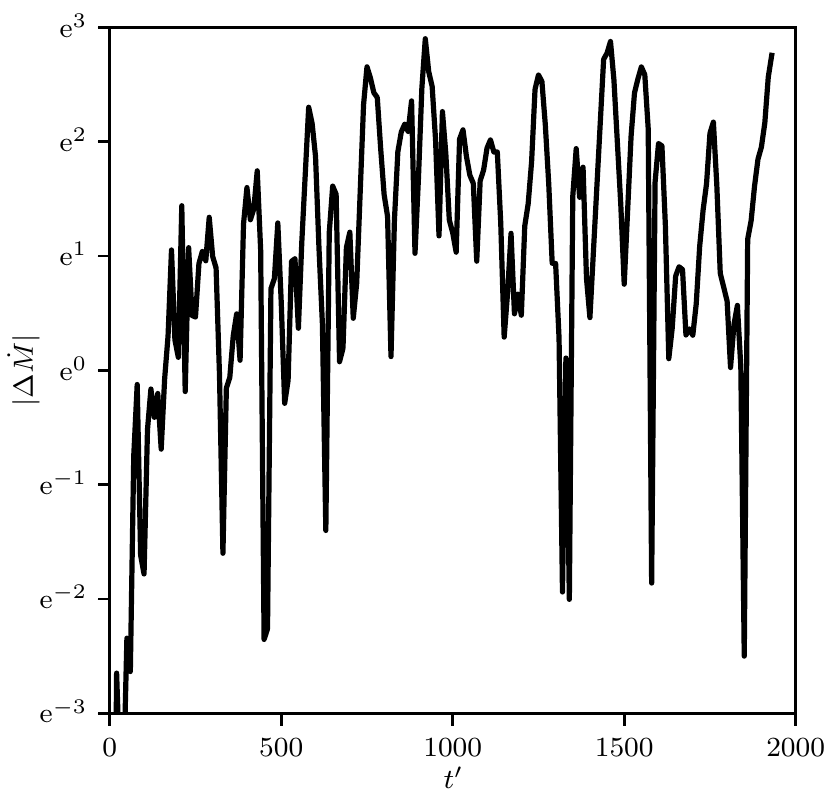}
  \caption{Absolute difference in accretion rates between the Level~0 and Level~2 simulations as a function of time. We expect exponential growth at first, followed by saturation. The slope of the growth phase $0 < t' \lesssim 250$ corresponds to the Lyapunov exponent. Saturation occurs for $t' \gtrsim 500$, justifying our choice of time interval to use in our analyses. \label{fig:lyapunov}}
\end{figure}

Finally, we consider the characteristic Lyapunov time for chaotic systems to diverge from one another. Figure~\ref{fig:lyapunov} shows the difference between the Level~0 and Level~2 $\mdot$ curves from Figure~\ref{fig:accretion}. Starting at $t' = 0$, we expect the difference to roughly diverge exponentially, before eventually saturating at some level. From the figure, we see four $\mathrm{e}$-foldings occur in the first approximately $250$ gravitational times, and the difference saturates by $t' \approx 500$. That is, by $t' = 500$ multiple Lyapunov timescales have elapsed, and the simulations are largely no longer diverging in their characteristics. The difference plots between other pairs of resolutions look similar.

\subsection{Interpretation of Results}

Our three grids are identical in setup except for resolution in the jet region. At the base of the jet, they have $8$, $12$, and $24$ cells from $\theta = 0$ to $\theta = 3 \pi / 16$ ($4$, $8$, and $16$ cells from $\theta = 0$ to $\theta = \pi / 8$). In azimuth they have $64$, $128$, and $256$ cells immediately surrounding the polar axis. For most quantities of interest these simulations produce similar results, though there are some small discrepancies.

Given the somewhat confined nature of the jet -- the high-velocity outflow is close to the axes and away from the midplane containing the bulk of the infalling matter -- we expect the accretion flow to be unaffected by jet resolution. Indeed the average accretion rate is the same in all three cases, as is the average amount of magnetic flux kept near the horizon. The accretion rate displays slightly more time variability at the highest resolution, indicating there is some feedback from the base of the jet to the innermost parts of the accretion flow. Though capturing short-timescale variations requires sufficient nearby resolution, the averages are set by the supply of material at large radii. Thus we see no reason to doubt the average accretion properties of similar systems in the literature, even in cases with very low resolutions at the poles.

We expect the energy fluxes in the jet to be unaffected by resolution for a different reason. Conservative codes should, like nature, neither create nor destroy energy. (Numerical floors of course violate this precept, but we do not expect their effect to scale with resolution, as discussed in Section~\ref{sec:results:jet_power}.) There is only the chance that energy will be directed in the wrong way, but we see the same jetted outflow at all resolutions when fed by the same highly resolved thick accretion flow. That is, even with very few cells at the base of the jet, we see neither spurious advection of energy into the black hole nor significantly different amounts of the Blandford--Znajek process, relative to the case with many cells at the base.

Even if there is agreement on total energy, it might be the case that the energy is present in different forms. In particular, at low resolutions the magnetic field may undergo excessive numerical reconnection, converting into gas thermal energy. Figure~\ref{fig:jet_power_radius} shows this is not the case. While \citet{White2019} showed that resolution is important for capturing the correct magnetic behaviour in the disc, the flow in the jet is more steady and is in a uniform direction. Numerical reconnection occurs when opposing fields are brought into the same cell by advection;\ when the field and velocity are aligned, there is little chance for this to occur.

It is worth noting that the highly magnetized, high-velocity flow in the strongly curved spacetime near the horizon is especially prone to numerical failures. These failures manifest during variable inversion, when the code tries to find the primitive quantities ($\rho$, $\pgas$, and gas velocity) associated with the conserved quantities in a cell. The problem has proven important enough to warrant study in its own right \citep{Noble2006,Marti2015,Siegel2018,Ripperda2019,Kastaun2020}. There are occasional failures here, where a primitive state cannot be found and the code falls back to heuristics to replace the values in the cell. The agreement we see indicates this issue does not scale considerably with resolution, and that violating energy conservation is probably only a small effect here.

Discrepancies between the three simulations are seen when we look at the structure of the jet in more detail. Figure~\ref{fig:lateral} shows that while the velocity profile across the jet is in agreement, the magnetization levels are not. The differences in magnetization at larger radii can be traced back to differences at the very base of the jet, where the lower resolutions produce an over-magnetized core and a slightly under-magnetized region outside that.

This magnetization, however, may well be in considerable disagreement with nature. At the base of the jet there is always a stagnation surface, separating material that flows outward from material that falls into the black hole. This can be seen in Figure~\ref{fig:jet_power_radius} where the mass flux changes sign (between $r = 3.4$ and $r = 3.6$). The divergence in the velocity at this location is trying to evacuate the region, but finite-volume codes impose non-zero density floors to avoid the regime where the continuum equations they solve fail. As a result, mass is injected near the stagnation surface, artificially loading the jet. Physically, one expects pair creation to create a dilute plasma in the region, though such an effect is not incorporated into the vast majority of ideal MHD simulations of this type. Our goal here is not to solve this widespread shortcoming, but only to elucidate which properties of these simulations incorporating broadly assumed physics are resolved with typical grids. It may well be that future incorporation of additional physics will require resolution of smaller length scales.

Our investigation reveals a well-defined spiral structure present in all three jets, which can be seen in the location of the $\sigma$-weighted jet centre (Figure~\ref{fig:spiral}). Here we remain agnostic as to the exact cause. It may be the imprint of a spiral wave or Rayleigh--Taylor unstable bubbles in the midplane;\ alternatively it may be a manifestation of the kink instability to which such jets tend to be marginally unstable \citep{McKinney2006} and as was studied for example with similar simulations by \citet{Bromberg2016}. In any case, it is surprisingly well resolved even at the coarsest resolution, though the displacements from the axis are more well defined (less variable) at higher resolution. The agreement may be due to the fact that much of the coarsest part of the low-resolution grid (see Figure~\ref{fig:grid}) lies below the stagnation surface. The poorly-resolved material here has little chance to affect the jet at larger radii.

The newest class of observational tools used to probe black hole accretion is that of horizon-scale interferometry. We assess how jet resolution might affect observations similar to those of M87 done with the EHT \citep{EHT2019a,EHT2019b,EHT2019c,EHT2019d,EHT2019e,EHT2019f} by examining ray-traced images created from our simulations. When modelling M87 as it appears at $230\ \ghz$, there is some difference in the statistics of the light curves -- the amplitude of variability, as measured by the standard deviation of the set of flux values, increases with resolution. This agreement among the means but trend among the standard deviations persists when we look at $43\ \ghz$, lower than the EHT observations but useful for highlighting emission from the jet. It is also found when using a considerably different, edge-on viewing angle. While we find no evidence that ray tracing is grossly inaccurate as a result of low polar resolutions, caution should be exercised when interpreting quantitative details, such as variability statistics, of simulations performed with only a few cells across the jet.

It is possible that low-resolution simulations will fail to capture transient structure in the images. However, seeing such structure would be difficult given the effective resolution of EHT. While beyond the scope of this investigation, it may be possible to set a lower limit on acceptable simulation jet resolution based on an analysis of the impact it has on raw EHT data in the sparsely sampled Fourier plane, rather than in the well-sampled image planes we show here.

\subsection{Conclusions and Future Prospects}

Numerical studies such as this explore how well we can trust the results of simulations, which are a vital tool used to understand MAD accretion onto and relativistic jets launched from black holes. By focusing on resolution in the jet, we complement the disc resolution study of \citet{White2019}. Other complementary studies can be done. For example, a code comparison effort is underway (Olivares et al., in prep.) to determine if the different algorithmic details in codes like \athena{} produce comparable results for the same physical systems as examined here.

As far as jet resolution itself matters, we find that even relatively low resolutions produce similar results to an extremely expensive high-resolution simulation, as long as the disc is well resolved. There are slight differences in structure, especially in jet magnetization, though the inclusion of additional physics could well modify this structure. Variability, whether in accretion rate or modelled emitted light, also increases as resolution is added to the jet region.

Our analysis has focused on the effects of resolution in spherical coordinates, which dominate the literature for these physical systems. Some recent works -- \citet{Davelaar2019} and \citet{Ressler2020}, for example -- have used Cartesian coordinates (still with mesh refinement) to simulate MAD accretion, naturally capable of obtaining high resolution at the base of the jet without extremely small timesteps, though possibly at the expense of making angular momentum conservation more difficult. Future work could compare these two approaches in terms of their numerical properties.

\section*{Acknowledgements}

We thank O.~Blaes for useful discussions and suggestions. This research was supported in part by the National Science Foundation under grant NSF~PHY-1748958. This work used the Extreme Science and Engineering Discovery Environment (XSEDE) Stampede2 at the Texas Advanced Computing Center through allocation AST170012.

\section*{Data Availability}

The data underlying this article will be shared on reasonable request to the corresponding author. The codes used for simulation (\athena, \url{https://github.com/PrincetonUniversity/athena-public-version}) and ray tracing (\grtrans, \url{https://github.com/jadexter/grtrans}) are both publicly available.

\bibliographystyle{mnras}
\bibliography{references,references_extra}

\bsp
\label{lastpage}

\end{document}